\documentclass[12pt]{iopart}
\usepackage{iopams}

\begin{document}

\title[Scaling the extrinsic curvature...]{Scaling up the extrinsic curvature in asymptotically flat gravitational initial data:  Generating trapped surfaces }
\author{Shan Bai$^{1,2}$ and Niall \'O Murchadha$^1$}
\address{$^1$ Physics Department, University College Cork, Cork, Ireland}

 \address{$^2$ Institut de Math\'ematiques de Bourgogne, UMR 5584 CNRS, 9  Avenue Alain Savary, BP 47870, 21078 Dijon, France.}
\eads{\mailto{shan.bai@u-bourgogne.fr}, \mailto{niall@ucc.ie}}

\date{\today}

\begin{abstract}

  The existence of the initial value constraints means that specifying initial data for the Einstein equations is non-trivial. The standard method of constructing initial data in the asymptotically flat case is to choose an asymptotically flat 3-metric and a transverse-tracefree (TT) tensor on it. One can find a conformal transformation that maps these data into solutions of the constraints. In particular, the TT tensor becomes the extrinsic curvature of the 3-slice. We wish to understand how the physical solution changes as the free data is changed. In this paper we investigate an especially simple change: we multiply the TT tensor by a large constant. One might assume that this corresponds to pumping up the extrinsic curvature in the physical initial data. Unexpectedly,  we show that, while the conformal factor monotonically increases, the physical extrinsic curvature decreases. The increase in the conformal factor however means that the physical volume increases in such a way that the  ADM mass become unboundedly large.  In turn, the blow-up of the mass combined with the control we have on the extrinsic curvature allows us to show that trapped surfaces, i.e., surfaces that are simultaneously future and past trapped, appear in the physical initial data.
\end{abstract}


\maketitle

\section{Introduction}
When we consider General Relativity as a dynamical system, or try to construct solutions to the Einstein equations numerically, we draw on our understanding of and experience with other theories of physics. As in the case of  particle mechanics or electromagnetism, we can pose initial data, and use the field equations to propagate the system into the future (or even into the past if we wish). We need to give a Riemannian 3-metric $g_{ij}$, and a symmetric tensor $K^{ij}$ that is the extrinsic curvature of the 3-geometry. All regular solutions to the equations can be obtained in this way.

 Just as in electromagnetism, the initial data must satisfy constraints. This means that we must first specify `free' data, and then manipulate these data to map them into a solution of the constraints. What happens to the physical solution as one changes the free data? In this article we change the free data in a particularly simple way and investigate the consequences.

 Since electromagnetism is a massless spin-1 field theory, and gravity is a massless spin-2 field theory, it is not surprising that we can see parallels between them.  Consider the Maxwell free data as a pair of 3-vectors, $(\vec{A}, \vec{F})$. $\vec{A}$ is the magnetic vector potential, and $\vec{F}$ is the vector field from which we extract the electric field. Thus, the magnetic and electric fields can be generated by $\vec{B} = \vec{\nabla}\times\vec{A}$ and $\vec{E} = \vec{F} - \vec{\nabla} V$, where $V$ is a scalar function chosen to satisfy $\nabla^2V = \nabla_iF^i$.   Electromagnetism as a theory can be expressed in terms of $(\vec{A}, \vec{E})$ rather than the standard $(\vec{B}, \vec{E})$. We can see that $\vec{A}$ is analogous to the metric, since both are gauge dependent quantities, while $\vec{E}$ is the analogue of the extrinsic curvature, because $\vec{E}$ is essentially the time derivative of $\vec{A}$, and the extrinsic curvature is essentially the time derivative of the metric. Thus $\vec{F}$ is the analogue of the free data from which we extract the extrinsic curvature. It is clear that if we multiply $\vec{F}$ by any constant, then $\vec{E}$ will then be multiplied by the same amount. As a result, the electromagnetic energy density, $B^2 + E^2$, increases without limit, and then the total energy will also blow up. In this article we want to multiply the free data, which specifies the extrinsic curvature, by a large constant in order to see what happens. We will again show that the total gravitational energy increases without limit.

Initial data for the Einstein equations consists of two parts: (1) a 3-dimensional spacelike slice with a Riemannian metric, $\bar{g}_{ij}$, and (2) a symmetric 3-tensor $\bar{K}^{ij}$. The spacelike slice is to be regarded as a slice through a 4-dimensional pseudo-Riemannian manifold that satisfies the Einstein equations, and $\bar{K}^{ij}$ is the extrinsic curvature of the 3-surface embedded in the 4-manifold. This means that $\bar{K}^{ij}$ is the Lie derivative of the 3-metric along the timelike normal to the slice. These initial data cannot be freely specified; rather, they satisfy 4 constraints, which, in vacuum, read
\begin{eqnarray}
^{(3)}\bar{R} - \bar{K}^{ij}\bar{K}_{ij} + \bar{K}^2 &=& 0 \label{ham}\\
\bar{\nabla}^i(\bar{K}_{ij} - \bar{g}_{ij}\bar{K}) = 0, \label{mom}
\end{eqnarray}
where $^{(3)}\bar{R}$ is the scalar curvature of the 3-metric, $\bar{K} = \bar{g}_{ab}\bar{K}^{ab}$ is the trace of the extrinsic curvature, and $\bar{\nabla}^i$ is the covariant derivative with respect to $\bar{g}_{ij}$. Eq.(\ref{ham}) is called the Hamiltonian constraint, and Eq.(\ref{mom}) is called the momentum constraint. A comprehensive account of the constraints can be found in \cite{CB}, especially in Chapter VII; the classic article is \cite{ADM}. If $\bar{K} = 0$, the momentum constraint shows that $\bar{K^{ij}}$ is both divergence-free and trace-free. A symmetric 2-tensor is called a TT tensor if it is simultaneously divergence-free and trace-free.  It turns out that TT tensors are conformally covariant in the sense that if $F^{TT}_{ij}$ is TT with respect to a metric $g_{ij}$, then $\phi^{-2}F^{TT}_{ij}$ is TT with respect to $\bar{g}_{ij} = \phi^4 g_{ij}$.

The standard way of finding sets $(\bar{g}_{ij}, \bar{K}^{ij})$ that solve the constraints is the so-called conformal method which was initiated by Andr\'e Lichnerowicz \cite{L}. The idea is to specify a base-metric $g_{ij}$ and a TT tensor. The base-metric is conformally transformed via $\bar{g}_{ij} = \phi^4 g_{ij}$, choosing the conformal factor $\phi$ so as to satisfy simultaneously the Hamiltonian constraint and the momentum constraint.

This article deals specifically with constructing initial data that are asymptotically flat. This means that we choose the base metric to be asymptotically flat, and seek a conformal factor, $\phi$, such that $\phi \rightarrow 1$ at infinity to maintain asymptotic flatness. As discussed above, we will restrict our attention to `maximal' slices, initial data for which $K = 0$. This is a non-trivial assumption: there exist spacetimes which contain no maximal slice, see, for example, \cite{HDW}. On the other hand, there exist many spacetimes that do possess maximal slices. The maximal assumption is useful because it makes it easy to specify the `free' data.

In Section II, we will see what happens when we multiply the `free' TT tensor by a large constant. This maintains TT-ness of the tensor, and we can always find a suitable conformal factor to solve the constraints. We will show that, despite the fact that we continue to demand $\phi \rightarrow 1$ at infinity, the conformal factor monotonically blows up as we increase the TT tensor. Further, we can show that the conformal factor behaves like $\phi \approx 1 + E/2r + \dots$ at infinity, where $E$ is the ADM energy of the solution. In Section III we will show that $E$ increases monotonically so that the energy becomes unboundedly large. In Section IV we show that trapped surfaces must appear in the physical data when the scaling is large.

This paper is a companion to \cite{BaOM}, in which we did the same analysis on a compact manifold. In \cite{BaOM} we found that the behaviour of the conformal factor was complicated; we got blow-up but not uniform blow-up.  That paper therefore used a combination of numerical and analytic techniques to describe how the conformal factor changed. We assumed we would have similar difficulties in this calculation and did some numerical modeling. However,  on viewing the numerics it was clear that the the conformal factor behaved in a much more regular fashion: we clearly had uniform blow-up. This inspired us to go back and look again at the analytical work.  We were able to produce exact mathematical proofs of everything we wanted, and so we were able to eliminate the numerics. This, therefore, is a paper that has passed through a numerical phase and emerged finally as a purely analytic document.

\section{Scaling the Extrinsic Curvature}
The Hamiltonian constraint is
\begin{equation}
^{(3)}\bar{R} - \bar{K}^{ij}\bar{K}_{ij} + \bar{K}^2 = 0.
\end{equation}
We assume that we are given a suitably asymptotically flat metric, $g_{ij}$,  and a suitably asymptotically flat $TT$ tensor,  $K^{ij}_{TT}$, with respect to the given metric. We map this set of initial data to an asymptotically flat solution of the Hamiltonian constraint by using $\bar{g_{ij}} = \phi^4g_{ij}$, and $\bar{K^{ij}_{TT}} = \phi^{-10}K^{ij}_{TT}$, and finding an appropriate $\phi$. Since the extrinsic curvature stays $TT$ under this transformation, the momentum constraint is automatically satisfied. This reduces to  solving the Lichnerowicz equation
 \begin{equation}
\nabla^2\phi -\frac{R}{8}\phi + \frac{1}{8}  A^2 \phi^{-7} = 0\label{L}
\end{equation}
where $A^2= K^{TT}_{ij}K^{ij}_{TT}$. We impose the boundary condition $\phi \rightarrow 1$ at infinity.

If we have a maximal solution, i.e., $\bar{K} = 0$, then the Hamiltonian constraint simplifies slightly to $^{(3)}\bar{R} - \bar{K}^{ij}\bar{K}_{ij} = 0$, and an immediate consequence is $^{(3)}\bar{R} \ge 0$. If we have a solution to the Lichnerowicz equation Eq.(\ref{L}) we will conformally transform the base metric into one with non-negative scalar curvature. It turns out that there is an obstruction to this.  Riemannian 3-manifolds split into three classes, called the Yamabe classes \cite{Y}. Only metrics in the positive Yamabe class can be conformally transformed into metrics with non-negative scalar curvature. These are the metrics that can be conformally mapped into a closed, without boundary, compact manifold with constant positive scalar curvature. All we need is that the AF metric be conformally flat at least to order $1/r^2$. Obviously, if we want to construct a maximal solution to the Hamiltonian constraint via a conformal transformation, the base metric must be in the positive Yamabe class.  One can show that Eq.(\ref{L}) will have a unique positive solution if and only if the metric belongs to the positive Yamabe class \cite{OMY}.

We want to change the initial data and see what happens. In particular, we rescale the background TT tensor. We change $K^{ij}_{TT}$  to  $\alpha^4K^{ij}_{TT}$ where $\alpha$ is a constant. The new $K^{ij}$ is still TT and we continue to be able to solve the Lichnerowicz equation.  In this article we are interested in  the behaviour of the conformal factor as $\alpha \rightarrow \infty$.  This means that we write the Lichnerowicz equation as
 \begin{equation}
\nabla^2\phi -\frac{R}{8}\phi + \frac{1}{8} \alpha^8 A^2 \phi^{-7} = 0.\label{L1}
\end{equation}
We are, of course, particularly interested in the properties of the physical initial data, $(\bar{g}_{ij}, \bar{K}^{ij}_{TT}) = (\phi^4g_{ij}, \phi^{-10}\alpha^4K^{ij}_{TT})$, as $\alpha$ becomes large.  Since we are in the positive Yamabe class, we can always make a preliminary conformal transformation to map the metric to one which satisfies $R \equiv 0$. This conformal transformation depends only on the base metric, i.e., it is independent of $\alpha$. This reduces the equation we wish to analyse to
 \begin{equation}
\nabla^2\phi  + \frac{1}{8} \alpha^8 A^2 \phi^{-7} = 0. \label{L2}
\end{equation}
with, of course, $\phi \rightarrow 1$. We know that this equation will have a regular positive solution for any finite $\alpha$. What happens to $\phi$ as $\alpha$ becomes large? This equation would make no sense if $\phi$ remained regular and bounded as $\alpha \rightarrow \infty$, because we would have a finite term equalling a term that becomes unboundedly large. Therefore we expect that $\phi$ blows up as $\alpha$ becomes large. More precisely, we expect that $\phi$ blows up linearly with $\alpha$. This is why we choose $\alpha^4$ as the rescaling factor. We now introduce a normalized $\phi$, $\hat\phi = \phi/\alpha$. This allows us to rewrite Eq.(\ref{L2}) as
\begin{equation}
\nabla^2\hat\phi  + \frac{1}{8}  A^2\hat \phi^{-7} = 0, \qquad \hat\phi \rightarrow 1/\alpha\ \  {\rm at\ \ } \infty. \label{L3}
\end{equation}
Obviously, $\hat\phi$ depends on $\alpha$. Let us consider a slightly different equation,
\begin{equation}
\nabla^2\psi  + \frac{1}{8}  A^2\psi^{-7} = 0, \qquad \psi  \rightarrow 0\ \  {\rm at\ \ } \infty. \label{L4}
\end{equation}
This equation has a regular positive solution which is unique.  This $\psi$, when used as a conformal factor, results in a compact without boundary manifold satisfying $R = \psi^{-12}A^2$. We can do this in two stages. First:  the  Yamabe theorem \cite{Y} tells  us that we can conformally map the asymptotically flat metric to a compact manifold of constant positive scalar curvature, $R_0$.  Second: following this transformation, Eq.(\ref{L4}) becomes
\begin{equation}
\nabla^2\phi -\frac{R_0}{8}\phi + \frac{1}{8}  \hat{A}^2 \phi^{-7} = 0.\label{L5}
\end{equation}
This has a regular solution on the compact manifold. The solution of Eq.(\ref{L4}) is just the product of the compactifying conformal factor by the solution of Eq.(\ref{L5}). Note: this requires that $\hat{A}^2$ be well behaved on the compact manifold. This is obviously satisfied if $A^2$ falls off like $r^{-12}$ or faster on the asymptotically flat manifold.

The key reason for introducing $\psi$ is that we will show
\begin{equation}
\hat\phi \rightarrow \psi\ \ {\rm as}\ \ \alpha \rightarrow \infty. \label{L6}
\end{equation}

{\bf First}: we can show that $\hat\phi$ monotonically decreases as $\alpha$ increases.  To see this, differentiate Eq.(\ref{L3}) by $\alpha$ to give
\begin{equation}
\nabla^2\frac{d\hat\phi}{d\alpha}  - \frac{7}{8}  A^2\hat\phi^{-8} \frac{d\hat\phi}{d\alpha} = 0, \qquad \frac{d\hat\phi}{d\alpha}  \rightarrow -1/\alpha^2\ \  {\rm at\ \ } \infty. \label{d1}
\end{equation}
It is clear that $d\hat\phi/d\alpha$ is negative at infinity. Let us assume that it is positive somewhere in the interior. If it is, then $d\hat\phi/d\alpha$ will have a positive maximum.  At such a point, both of the terms in Eq.(\ref{d1}) are negative, and this cannot be the case. Therefore we have $d\hat\phi/d\alpha < 0$. This means that $\hat\phi$ monotonically decreases as $\alpha$ increases.

{\bf Second}: we can show that $\hat\phi - 1/\alpha$ monotonically increases as $\alpha$ increases. The equation that $ \hat\phi' = \hat\phi - 1/\alpha$ satisfies is
\begin{equation}
\nabla^2\hat\phi'  + \frac{1}{8}  A^2(\hat \phi' + 1/\alpha)^{-7} = 0, \qquad \hat\phi' \rightarrow 0\ \  {\rm at\ \ } \infty. \label{L7}
\end{equation}
Again, we differentiate this with respect to $\alpha$ and get
\begin{equation}
 \nabla^2\frac{d\hat\phi'}{d\alpha}  - \frac{7}{8}  A^2(\hat\phi' + 1/\alpha)^{-8} \frac{d\hat\phi'}{d\alpha}   + \frac{7}{8\alpha^2}  A^2(\hat\phi' + 1/\alpha)^{-8} =  0,\qquad \frac{d\hat\phi'}{d\alpha}  \rightarrow 0\ \  {\rm at\ \ } \infty. \label{d2}
\end{equation}
Let us assume that $d\hat\phi'/d\alpha$ is negative somewhere in the interior. Then there must exist a point where it is a negative minimum. At such a point all the three terms in Eq.(\ref{d2}) are positive which is an obvious contradiction. Therefore we have $d\hat\phi'/d\alpha > 0$. This means that $\hat\phi' = \hat\phi - 1/\alpha$ monotonically increases as $\alpha$ increases.

Furthermore, we can also show that $\psi$ satisfies $\hat\phi > \psi > \hat\phi'$. To see this, let us first subtract Eq.(\ref{L4}) from Eq.(\ref{L3}) to get
\begin{equation}
\nabla^2(\hat\phi  - \psi)+ \frac{1}{8}  A^2(\hat \phi^{-7} - \psi^{-7})= 0,\qquad \hat\phi - \psi \rightarrow 1/\alpha\ \  {\rm at\ \ } \infty. \label{L8}
\end{equation}
Let us assume that $(\hat\phi  - \psi)$ goes negative in the interior. This means that there will be a negative minimum. At this point we have $\nabla^2(\hat\phi  - \psi) \ge 0$ and, since $\hat\phi  < \psi $ we have $\hat\phi^{-7} >   \psi^{-7}$. Therefore both terms in Eq.(\ref{L8}) are positive which  cannot be the case. So we must have $\hat\phi  - \psi > 0$ for all $\alpha$.

To show that $\psi > \hat\phi'$, we need to subtract Eq.(\ref{L7}) from Eq.(\ref{L4}) to get
\begin{equation}
\nabla^2(\psi  - \hat\phi')+ \frac{1}{8}  A^2[\psi^{-7} - (\hat\phi' + 1/\alpha)^{-7}] = 0,\qquad   (\psi - \hat\phi') \rightarrow 0\ \  {\rm at\ \ } \infty. \label{L9}
\end{equation}
Let us assume that we have a region with $\psi  < \hat\phi'$. In such a region $\psi^{-7} > \hat\phi'^{-7} > (\hat\phi' + 1/\alpha)^{-7}$. In this region $\psi  - \hat\phi'$ will have a local minimum. At that point both terms  in Eq.(\ref{L9}) will be positive, again a contradiction. This implies $\psi  > \hat\phi'$.

Of course, we have $\hat\phi  - \hat\phi' = 1/\alpha$. This gap goes uniformly to zero as $\alpha \rightarrow \infty$ and we have $\hat\phi > \psi  > \hat\phi'$. Therefore $\hat\phi$ and $\hat\phi'$ uniformly approach $\psi$, one from above and one from below, as $\alpha \rightarrow \infty$.

\section{Controlling the ADM mass}

We know that the leading term of $\psi$ is of the form $C/2r$ and we can easily show
\begin{eqnarray}
C = -\frac{1}{2\pi}\oint_{\infty}\nabla_i\psi dS^i &=& - \frac{1}{2\pi}\int \nabla^2 \psi dv \cr&=& \frac{1}{16\pi} \int A^2\psi^{-7}dv. \label{C}
\end{eqnarray}
This means that $C$ is positive and finite.

 Now $\hat\phi$ will behave asymptotically like $D/2r + 1/\alpha$ with $D$ depending on $\alpha$. We will show that $D \rightarrow C$ as $\alpha \rightarrow \infty$. We have that
  \begin{equation}
D = -\frac{1}{2\pi}\oint_{\infty}\nabla_i\hat\phi dS^i = - \frac{1}{2\pi}\int \nabla^2 \hat\phi dv = \frac{1}{16\pi}\int A^2\hat\phi^{-7}dv. \label{D}
\end{equation}
We know that $\hat\phi$ is bigger than $\psi$ but less than  $\psi + 1/\alpha$. Thus
\begin{equation}
\psi < \hat\phi < \psi + 1/\alpha.
\end{equation}
Consequently
\begin{equation}
(\psi + 1/\alpha)^{-7} < \hat\phi^{-7} < \psi^{-7},\label{psi}
\end{equation}
which implies
\begin{equation}
\frac{1}{16\pi} \int A^2(\psi + 1/\alpha)^{-7}dv < \frac{1}{16\pi} \int A^2\hat\phi^{-7} dv
< \frac{1}{16\pi} \int A^2\psi^{-7}dv.
\end{equation}
This can be immediately rewritten as
\begin{equation}
\frac{1}{16\pi} \int A^2(\psi + 1/\alpha)^{-7}dv < D < C. \label{CD}
\end{equation}
Using the monotone convergence theorem we see that $\frac{1}{16\pi} \int A^2(\psi + 1/\alpha)^{-7}dv \rightarrow C$ as $\alpha \rightarrow \infty$. Therefore we get $D \rightarrow C$.
 From this we conclude that the coefficient of the $1/2r$ part of the `physical' conformal factor, $\phi = \alpha \hat\phi$, behaves like $\alpha D \approx \alpha C$. In other words, it blows up linearly with $\alpha$.

 The energy of the physical metric is the sum of the monopole part of the `background' which is finite and bounded plus the monopole part of the conformal factor which is positive and becomes large. Thus the total energy is positive and increases monotonically with $\alpha$. This, by the way, is strong-field positive energy proof.

The conformal factor which maps to a physical asymptotically flat data set is $\phi$, the solution of Eq.(\ref{L2}).  We use this to generate a solution of the constraints $(\bar{g}_{ij}, \bar{K}^{ij}_{TT}) = (\phi^4 g_{ij}, \phi^{-10}\alpha^4K^{ij}_{TT})$. From this we get  $\bar{K}^{ij}_{TT}\bar{K}^{TT}_{ij} = \phi^{-12}\alpha^8K^{ij}_{TT}K^{TT}_{ij}$. Since $\phi$ scales with $\alpha$, this goes pointwise to zero like $\alpha^{-4}$.

The physics here is a bit tricky. We can think of $\bar{K}^{ij}_{TT}\bar{K}^{TT}_{ij}$ as the analogue of $E^2$ in electromagnetism, and so can be thought of as the kinetic energy part of the gravitational wave energy density. This shrinks like $\alpha^{-4}$. However, the physical volume increases as $\alpha^6$. This is why we see the total energy increasing despite the decrease of the energy density.

Nevertheless, it is clear that the physical extrinsic curvature shrinks as $\alpha$ increases. Therefore the physical solution looks more and more like a moment of time symmetry data set. Each physical solution will have an ADM energy-momentum. Since we assume the extrinsic curvature falls off faster than $1/r^2$, the ADM momentum vanishes, and the ADM energy becomes the ADM mass. This will be contained in the $1/r$ part of the physical metric. This will be made up of two parts. One part will be the `mass'  contribution of the background metric, $g_{ij}$, which is a fixed number. The other part is the coefficient of the $1/2r$ term in the conformal factor, $\phi$. This is $\alpha D \approx \alpha C$. Therefore, for large $\alpha$, the ADM mass diverges linearly with $\alpha$. This is a `strong-field' positive energy proof since we show that the energy is positive for large $\alpha$ without any restriction on the background energy.

Further, we can multiply Eq.(\ref{psi}) by $\alpha$ to see that $\phi$ satisfies
\begin{equation}
\alpha\psi < \phi < \alpha\psi +1. \label{phi}
\end{equation}
Therefore the conformal factor blows up uniformly and linearly with $\alpha$.

\section{Appearance of trapped surfaces}
Trapped surfaces are defined by negative outgoing null expansions.  Consider a closed 2-surface in an initial data set, with outgoing unit spatial normal, $n^i$. The outgoing null expansions are defined by
\begin{equation}\label{TS}
H_{\pm} = k \pm (\bar{g}_{ij} - n_in_j)\bar{K}^{ij} = \nabla_in^i \pm (\bar{g}_{ij} - n_in_j)\bar{K}^{ij}.
\end{equation}

Given a moment-of-time-symmetry slice in isotropic coordinates through the Schwarzschild spacetime, we know that the sphere defined by $r = m/2$ is the apparent horizon and all surfaces with $r < m/2$ are trapped. In particular, we can show that the mean curvature, $k$, of the surface defined by $r = m/4$ equals $-2/r = -8/m$.  We have a sequence of physical metrics, labeled by $\alpha$. On these metrics we can introduce quasi-isotropic coordinates near infinity, and the surfaces labeled by $r = m/4$ will be included in these domains for large enough $m$. We work out the null expansion of these surfaces. This will have a negative leading term, $-8/m$. Now we show that the corrections all fall off faster than $1/r$, i.e., faster than $1/m$. Thus, for large $m$, the leading negative term dominates and the surfaces are trapped.

The key term in our analysis is $k = \nabla_in^i$, the mean curvature of the 2-surface in the 3-space. We can ignore the extrinsic curvature term in this calculation because it falls off rapidly both with radius and with $\alpha$. Therefore to find a trapped surface we need only find a 2-surface with negative mean curvature, $k$. This already shows that we are getting trapped surfaces in the language of Penrose. The surfaces will be simultaneously future and past trapped.

To do this we need to understand the behaviour of $\phi$ at large radii. From Eq.(\ref{phi}) we know that we can write $\phi$ as
\begin{equation}
\phi = 1 + \frac{\alpha D}{2r} + f,\label{f}
\end{equation}
where $f$ is dominated by the dipole moment, so it falls off like $1/r^2$, and grows proportional to $\alpha$.  This means $f = O(\alpha/r^2)$. We also need control of the gradient of $\phi$, i.e., the gradient of $f$. To do this, let us return to Eq.(\ref{L3}) and differentiate it with respect to, say, $x$ to get
\begin{equation}
\nabla^2\frac{d\hat\phi}{dx}  - \frac{7}{8}  A^2\hat \phi^{-8}\frac{d\hat\phi}{dx} = S, \qquad \frac{d\hat\phi}{dx} \rightarrow 0\ \  {\rm at\ \ } \infty. \label{L10}
\end{equation}
where $S$ is a source term which comes from differentiating $A^2$ and the metric. This is a nice linear elliptic equation which guarantees that $d\hat\phi/dx$ is well behaved. Since $\phi \approx \alpha\hat{\phi}$, we can deduce that $df/dx = O(\alpha/r^3)$.

Let us assume that the base metric is conformally flat outside some region of compact support and let us compute the mean curvature of a spherical surface $r = r_0$, assuming that the physical metric is conformally flat, $\bar{g}_{ij} = \phi^4\delta_{ij}$.
 The flat
space unit normal, $q_i = (x/r, y/r, z/r)$ is  proportional
to the normal in the physical space. We define $|q|^2 =
\bar{g}^{ij}q_iq_j$. The unit normal to the surface is then $n_i =
q_i/|q|$. We can now compute
 \begin{eqnarray}
k &=& \nabla_in^i = \frac{1}{\sqrt{\bar{g}}}\partial_i(\sqrt{\bar{g}}n^i) = \frac{1}{\sqrt{\bar{g}}}\partial_i(\sqrt{\bar{g}}\bar{g}^{im}n_m)\cr &=&
\frac{1}{(1 + m_t/2r +f)^6}\partial_i[(1 + \alpha D/2r + f)^4(x/r, y/r,
z/r)]\cr &=& \frac{2(1 - \alpha D/2r +f + x^i\partial_if/2)}{r(1 + \alpha D/2r
+ f)^3}.\label{k}
\end{eqnarray}
The mean curvature will be negative if
\begin{equation}
1 - \alpha D/2r +f + x^i\partial_if/2 < 0.\label{k1}
\end{equation}
We know that
\begin{equation}
f + x^i\partial_if/2 < C_3\alpha/r^{2}
\end{equation}
where $C_3$ is a constant independent of $\alpha$.
If we choose $r = m/4 = \alpha D/4$, we get
\begin{equation}
1 - \alpha D/2r +f + x^i\partial_if/2 <  1 - 2 + \frac{16C_3}{\alpha D}.\label{k2}
\end{equation}
The third term will be less than 1 for $\alpha$ large and so the mean
curvature goes negative.

 It is a straightforward exercise to show that if the base metric is not conformally flat, but is well behaved near infinity, the correction to $k$ falls of quickly. In the same way, as we have indicated above, the correction due to the extrinsic curvature is also negligible.

 This article is closely related to those cited in \cite{BOM}. The space of free data for the maximal constraints consists of smooth, Riemannian 3-metrics that have positive Yamabe constant together with finite TT tensors. We want to find out what happens as we approach the boundary of this space. In the articles cited in \cite{BOM}  we kept everything regular as we looked at a sequence of metrics along which the Yamabe constant went to zero. In this article we again keep everything regular but let the TT tensor diverge. This is a very different part of the boundary of the space of free data. Nevertheless, we find very similar behaviour. The conformal factor blows up, the ADM mass diverges to $+\infty$ and  horizons appear.

  Unfortunately, this is not enough to construct a valid positive energy proof because one needs to investigate the other parts of the boundary, for example we need to consider sequences of metrics which either become `rough' or cease to be Riemannian.

   \section{Appendix: Showing $D \rightarrow C$}

 It is `obvious' that as $\alpha \rightarrow \infty,\ \ \frac{1}{16\pi} \int A^2(\psi + 1/\alpha)^{-7}dv \rightarrow C$, but we have difficulty proving it.  Here we will prove something weaker, but this result is really all we need.

Pick a  radius $r = r_0$ such that
\begin{equation}
\frac{1}{16\pi} \int_{B(r_0)} A^2\psi^{-7} dv  > \frac{C}{2}\label{B}
\end{equation}
where $B(r_0)$ is the ball inside $r = r_0$. We know that both $\psi$ and $r_0$ are independent of $\alpha$. Since $\psi$ is positive so it will have a minimum value inside $B(r_0)$. We pick $\alpha_0$ so that $1/\alpha_0$ equals this minimum value. Therefore, for all $\alpha > \alpha_0$, we have that, for $r < r_0,  \ \ \psi > 1/\alpha$.  Therefore for $r < r_0$ we have $\psi + 1/\alpha < 2\psi$. In turn we get
 \begin{eqnarray}
\frac{1}{16\pi} \int A^2(\psi + 1/\alpha)^{-7}dv& > &\frac{1}{16\pi} \int_{B(r_0)} A^2(\psi + 1/\alpha)^{-7}dv\cr
>  \frac{1}{16\pi} \int_{B(r_0)} A^2(2\psi )^{-7}dv& = &\frac{1}{2^{11}\pi} \int_{B(r_0)} A^2(\psi )^{-7}dv
> \frac {C}{2^8}. \label{C'}
\end{eqnarray}
This means that we can write Eq.(\ref{CD}) as
\begin{equation}
\frac {C}{2^8} < D < C.\label{CD2}
\end{equation}
We can adjust this proof so as to replace the  $2^8$ by a number which is as close to unity as we wish. Pick a  $\beta << 1$. We move $r_0$ out until we get
\begin{equation}
\frac{1}{16\pi} \int_{B(r_0)} A^2\psi^{-7} dv  > \frac{C}{1 + \beta}.\label{B'}
\end{equation}
We also increase $\alpha_0$ such that for all $\alpha > \alpha_0$ we have that, for $r < r_0,  \ \ \beta\psi > 1/\alpha$.  Therefore for $r < r_0$ we have $\psi + 1/\alpha < (1 + \beta)\psi$. Then we get
\begin{equation}
\frac {C}{(1 + \beta)^8} < D < C.\label{CD3}
\end{equation}
This equation now holds for all $\alpha > \alpha_0$. In other words, we can make $D$ as close to $C$ as we wish by picking a large $\alpha$.

\ack
We thank the anonymous referee for pointing out the use of the monotone convergence theorem to us.
SB and N\'OM were supported by  Grant 07/RFP/PHYF148 from Science Foundation
Ireland. SB is also supported by a grant from the state of Burgundy.


\section*{References}

\begin{thebibliography}{}
\bibitem{CB} Choquet-Bruhat, Y., {\it General Relativity and the Einstein Equations,} (Oxford, OUP, 2009).
\bibitem{ADM} Arnowitt, R., Deser, S., and Misner, C., in {\it Gravitation: an introduction to current research,} ed. L. Witten (Wiley, New York, 1962).
\bibitem{L} Lichnerowicz, A., J. Math. Pures Appl. {\bf 23}, 39 (1944).
\bibitem{HDW} D. H. Witt, ArXiv: 0908.3205 [qr-qc] (2009).
\bibitem{BaOM}Bai, S., and \'O Murchadha, N., Phys. Rev. {\bf D 85}, 044028 (2012).
\bibitem{B} Brill, D., {\it On spacetimes without maximal surfaces} in {\it Proceedings of the third Marcel Grossmann meeting} ed. Hu Ning (Science Press and North Holland 1983) pp. 79 -87; Witt, H.D., gr-qc 0908.3205 (2009).

\bibitem{Y}  Yamabe, H., Osaka Math J. {\bf 12}, 21 (1960);  Schoen, R.,
J. Diff. Geom. {\bf20}, 479 (1984); Lee, J., and Parker, T. Bull. Am. Math. Soc. {\bf 17}, 37 (1987).

\bibitem{OMY} \'O Murchadha, N., and York, J.W., J. Math. Phys. {\bf 14}, 1551 (1973).

\bibitem{BOM} Beig, R., and  \'O Murchadha, N., Phys. Rev. Lett.{\bf 66,} 2421 (1991);
Class. Quantum Grav. {\bf 64}, 419 (1994); Class. Quantum Grav. {\bf
13}, 739 (1996); \'O Murchadha, N. and Xie, N. to be published.

\end{thebibliography}
\end{document}